\documentclass[prb,twocolumn,superscriptaddress]{revtex4}
\usepackage{graphicx}
\usepackage{amsfonts,amsmath,amssymb,float}
\usepackage{bm,dsfont}
\usepackage{color}
\usepackage{bbm}
\usepackage{longtable}
\usepackage[dvips]{epsfig}
\usepackage{amsmath,amssymb,lscape,float}
\usepackage{hyperref}
\usepackage{listings}

\begin{document}
\title{Analog quantum simulation of small-polaron physics in arrays of neutral atoms
with Rydberg-dressed resonant dipole-dipole interaction}

\author{Vladimir M. Stojanovi\'c}
\email{vladimir.stojanovic@physik.tu-darmstadt.de}
\affiliation{Institut f\"{u}r Angewandte Physik, Technical
University of Darmstadt, D-64289 Darmstadt, Germany}

\date{\today}

\begin{abstract}
Recent years have seen growing interest in sharp polaronic transitions in systems with 
strongly momentum-dependent interactions of an itinerant excitation (electron, hole, exciton)
with dispersionless phonons. This work presents a scheme for investigating such phenomena in a controllable fashion within the framework of an analog quantum simulator based on an 
array of neutral atoms in optical tweezers. The envisioned analog simulator, in which the atoms 
interact through Rydberg-dressed resonant dipole-dipole interaction, allows one to study 
the rich interplay of Peierls- and breathing-mode-type excitation-phonon interactions. Based on a numerically-exact treatment of one special case of this system -- namely, the case with equal 
Peierls- and breathing-mode coupling strengths -- a sharp small-polaron transition was shown to 
take place for a critical value of the effective excitation-phonon coupling strength. This transition signifies the change from a completely bare (undressed by phonons) excitation below the transition 
point and a strongly phonon-dressed one (small polaron) above it. This work also highlights the comparative advantages of highly-controllable Rydberg-atom-based systems to other physical 
platforms for simulating polaronic phenomenology, which could be exploited to study the 
nonequilibrium dynamics of the small-polaron formation.
\end{abstract}

\maketitle

\section{Introduction}
Ensembles of Rydberg atoms~\cite{GallagherBOOK} have in recent years established 
themselves as a rich playground for analog simulation of quantum many-body systems~\cite{Tamura+:20,Browaeys+Lahaye:20}. While many interesting phenomena 
can be emulated with such ensembles 
confined in optical lattices~\cite{Hollerith+:19}, the potential of Rydberg-atom-based systems in the 
quantum-simulation context has reached new heights with the recent advances in scalability pertaining 
to arrays of individual dipole traps (optical-tweezer arrays)~\cite{Bernien+:17,Barredo+:18}. Owing to the possibility of integrating multiqubit storage, readout, and 
transport~\cite{Beugnon+:07} in these systems, as well as their inherent capability for the coherent 
control of spin- and motional states of trapped atoms~\cite{Kaufman+:12}, tweezer arrays have acquired 
their present status of the most powerful platform for analog quantum simulation with neutral atoms.

Interwoven with the phenomenon of Rydberg blockade, the off-resonant (van der Waals) dipole-dipole interaction 
in Rydberg ensembles allows the realization of the paradigmatic Ising model of coupled spins. At the same time, 
the resonant (long-ranged) dipole-dipole interaction naturally leads to the realization of the spin-$1/2$ $XY$
model, which in an alternative physical picture maps onto models describing excitation 
transport~\cite{Browaeys+Lahaye:20}. In turn, such resonant energy transfer 
due to dipole-dipole interaction lies at the heart of the phenomena such as exciton dynamics in molecular 
crystals or light-harvesting systems~\cite{Ates+:08}. Continuing on the established trail of 
quantum simulations of systems with progressively increasing level of complexity, the present paper adds 
an additional ingredient to this last research strand and focusses on the behavior of an itinerant excitation 
coupled to zero-dimensional bosonic degrees of freedom. More precisely, it considers a situation 
where those bosons mimic the behavior of dispersionless phonons in solid-state systems and the ensuing 
excitation-phonon (e-ph) interaction is strong enough to allow the formation of a heavily-dressed 
quasiparticle known as {\em small polaron}~\cite{Emin:82}.

The concept of small polarons emerged from studies of itinerant excitations (electron, hole, exciton) in narrow-band 
semiconductors or insulators, where such excitations interact with the host-crystal lattice vibrations -- quantized
into dispersionless (Einstein-like) phonons -- through short-ranged, non-polar e-ph interactions. Such physical 
circumstances are typically described within the framework of the Holstein molecular-crystal model~\cite{Holstein:59,Wellein+Fehske:97}, 
a lattice model that only takes into account purely local coupling of the excitation density on a given lattice 
site and the Einstein-phonon displacement on that same site. Yet, the need to capture subtle e-ph interaction 
effects -- akin to those encountered in real materials~\cite{Sio+:19} -- has prompted investigations of nonlocal 
e-ph interactions~\cite{Stojanovic+:04,Shneyder+:20,Hannewald++:04,Stojanovic++:10}. 

It is known that the Gerlach-L\"{o}wen theorem rules out the existence of nonanalyticities
in the ground-state-related quantities characterizing a single, phonon-dressed excitation 
for certain types of e-ph interactions (namely, those whose corresponding e-ph 
vertex function is either completely momentum-independent or depends on the 
phonon- but not on the bare-excitation quasimomentum). The aforementioned line of research into nonlocal e-ph-coupling mechanisms has, however, even led 
to the discovery of sharp small-polaron transitions~\cite{Stojanovic:08,Sous+:17,Stojanovic+:14} in models with strongly momentum-dependent e-ph interactions~\cite{Stojanovic:20}, with 
Peierls-type e-ph coupling being the primary example. 

In this paper, it is shown that optical-tweezer arrays of cold atoms with Rydberg-dressed resonant dipole-dipole
interaction allow demonstration of subtle small-polaron effects, including the occurrence of sharp transitions. The physical 
mechanism that enables this analog simulation is Rydberg dressing, which -- generally speaking -- allows one to induce 
very strong dipole-dipole interactions even among atoms that almost reside in their ground states, i.e., with only a
small admixture of Rydberg states resulting from an off-resonant-laser coupling between ground- and Rydberg states.
In particular, what makes dressed Rydberg states particularly appealing from the standpoint of analog quantum 
simulation are their significantly longer lifetimes compared to their ordinary Rydberg counterparts~\cite{Pupillo+:10}.

Complementing the analog simulators of SP physics based on trapped ions~\cite{Stojanovic+:12,Mezzacapo+:12}, cold polar 
molecules~\cite{Herrera+Krems:11,Herrera+:13}, and superconducting circuits~\cite{Mei+:13,Stojanovic+:14,Stojanovic+Salom:19}, 
those based on Rydberg atoms/ions have been discussed in the past~\cite{Hague+MacCormick}. However, the latter did not 
allow one to reach the strong-coupling regime where sharp small-polaron
transitions take place. In contrast to these previous studies, 
the present work shows that such transitions can be realized with
realistic values of relevant experimental parameters in a system 
of neutral atoms interacting through Rydberg-dressed resonant 
dipole-dipole interaction. 

The remainder of this work is organized as follows. In Sec.~\ref{AnalogSimul}
the central physical mechanisms behind the envisioned neutral-atom analog simulator -- Rydberg-dressed resonant dipole-dipole interaction -- is described in great detail. The following Sec.~\ref{EPHmodel} 
begins with the derivation of the effective single-excitation 
Hamiltonian of this system that describes the competition of Peierls-
and breathing-mode-type e-ph interactions. This is followed with 
a discussion of the special case of this Hamiltonian in which these two mechanisms have the same coupling strengths. The principal findings of
this work, with emphasis on sharp small-polaron transitions -- are presented 
and discussed in Sec.~\ref{ResDisc}. Finally, the paper is summarized in Sec.~\ref{SummConcl}.


\section{Neutral-atom analog simulator} \label{AnalogSimul}
\subsection{System and Rydberg-dressing mechanism}
\begin{figure}[t!]
\includegraphics[clip,width=8.25cm]{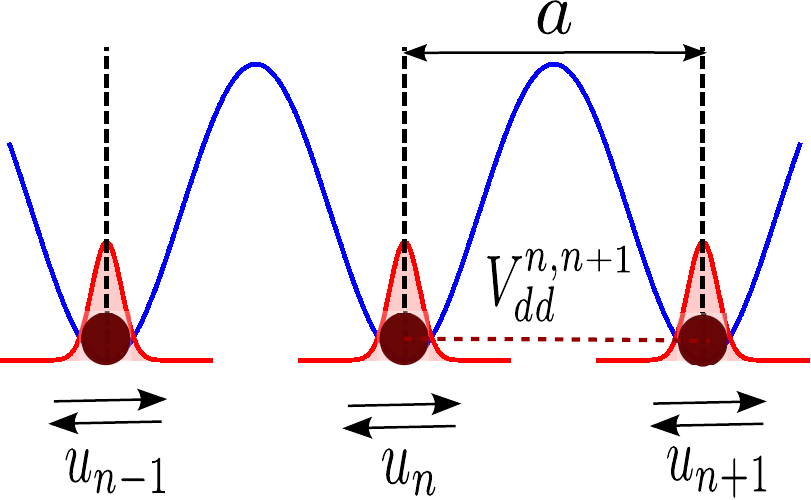}
\caption{\label{fig:system}A finite segment of the 
neutral-atom system under consideration, with three atoms confined in individual optical dipole traps; $a$ is the distance between the minima of adjacent traps. Assuming that adjacent atoms are in states corresponding to the 
same value of the principal quantum number $n_{\textrm{q}}$ but different values of the angular-momentum quantum number $l$, the atoms 
interact through resonant dipole-dipole interaction
(here denoted by $V^{n,n+1}_{dd}$ for atoms $n$ and $n+1$).}
\end{figure}

The envisioned analog simulator [pictorially illustrated 
in Fig.~\ref{fig:system}] entails a one-dimensional (1D) array of 
neutral atoms (mass $m$), for example $^{87}$Rb, each confined in its individual optical microtrap. These microtraps are assumed to be 
approximately harmonic, with distance $a$ between adjacent ones and trap frequency $\omega_{\textrm{tr}}$. 
Under the assumption of strong confinement, the vibrational motion of atoms in the vicinity of the minima of their respective microtraps 
can be quantized into dispersionless bosons that in the following will have the physical meaning of Einstein-like phonons; $b_n^{\dagger}$ 
($b_n$) creates (destroys) such a phonon with frequency $\omega_{\textrm{ph}}=\omega_{\textrm{tr}}$  in the $n$-th microtrap ($n=1,\ldots,N$). 
The displacement of atom $n$ from its equilibrium position is then given by $u_n\equiv(\hbar/2m\omega_{\textrm{ph}})^{1/2}(b_n+b_n^{\dagger})$. 
As a consequence of this vibrational motion, the interatomic distance dynamically fluctuates, so that, e.g., the distance between 
atoms $n$ and $n+1$ is given by $a+u_{n+1}-u_n$.

\begin{figure}[t!]
\includegraphics[clip,width=8.25cm]{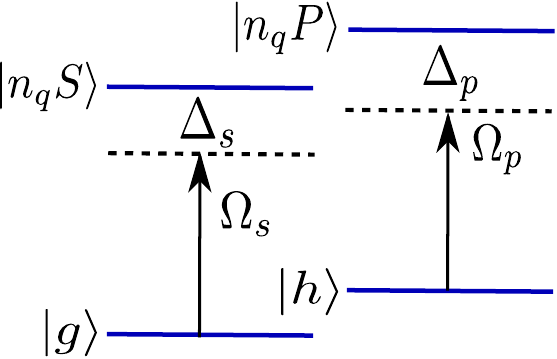}
\caption{\label{fig:EnergyDiag}Illustration of the 
Rydberg-dressed resonant dipole-dipole
interaction: two hyperfine sublevels of the atomic ground states (denoted by $|g\rangle$ and $|h\rangle$)
are off-resonantly coupled to a pair of highly excited Rydberg states ($|n_{\textrm q}S\rangle\equiv|n_{\textrm q},l=0\rangle$ and $|n_{\textrm q}P\rangle\equiv|n_{\textrm q},l=1\rangle$, respectively).}
\end{figure}

As usual for Rydberg-dressing of resonant dipole-dipole interactions, the envisioned simulator relies on the use of two laser couplings 
and four electronic states -- namely, two ground- and two high-lying Rydberg states -- of each atom~\cite{Wuester+:11}. Therefore, it requires off-resonant 
coherent coupling of two different ground states (i.e., two hyperfine levels of the atomic ground state) -- denoted in what follows as
$|g\rangle$ and $|h\rangle$ -- to a pair of highly excited Rydberg states $|n_{\textrm q}S\rangle\equiv|n_{\textrm q},l=0\rangle$ and 
$|n_{\textrm q}P\rangle\equiv|n_{\textrm q},l=1\rangle$ (for illustration, see 
Fig.~\ref{fig:EnergyDiag}), respectively, where $n_{\textrm q}$ is the principal- and 
$l$ the angular-momentum quantum number. Needless to say, this simple picture with four
relevant states per atom is valid provided that the detunings of both laser 
couplings are chosen such that all other transitions can be safely neglected. Besides, 
the required laser couplings involve in general two-photon (or multi-photon) transitions. 

Resulting from the envisaged Rydberg dressing, an atom resides in the dressed state $|0\rangle\approx|g\rangle +\alpha_s|n_{\textrm q}S\rangle$ 
or $|1\rangle\approx |h\rangle +\alpha_p|n_{\textrm q} P\rangle$), depending on which ground state ($|g\rangle$ or $|h\rangle$, respectively)
it occupied initially. Here $\alpha_{s,p}\equiv \Omega_{s,p}/(2\Delta_{s,p})$ are the effective dressing parameters, fixed by the total Rabi 
frequencies $\Omega_{s,p}$ of the driving fields and the total laser detunings $\Delta_{s,p}$. For simplicity, it is henceforth assumed that 
$\alpha_s=\alpha_p\equiv\alpha$, with $\alpha$ being the main control parameter in the system at hand.

The mechanism behind e-ph interaction in this system is coupling of the internal states of atoms to their motional degrees of 
freedom. The latter coupling is induced by dovetailing an additional ingredient -- namely, the aforementioned Rydberg dressing -- to the conventional 
mechanism of excitation transport originating from the long-range resonant dipole-dipole interaction (recall that this interaction scales 
as the inverse third power of the interatomic distance, i.e., $C_3 R^{-3}$)~\cite{Barredo+:15}. In the more conventional setting, 
with a pair of approximate two-level atoms with Rydberg states $|n_{\textrm q}S\rangle$ and $|n_{\textrm q}P\rangle$, the interaction manifests 
itself as the hopping of an $n_{\textrm q}P$ excitation among the pair of atoms, while in the system at hand the usual 
Rydberg states are supplanted by the dressed states $|0\rangle$ and $|1\rangle$~\cite{Wuester+:11}. The dynamical fluctuation of interatomic 
separations leads to an effective dependence of both the excitation's on-site energy and its hopping amplitude on the phonon degrees of freedom. 
This is the origin of two e-ph coupling terms reminiscent of those characteristic of solid-state 
systems~\cite{Stojanovic+:04,Shneyder+:20,Hannewald++:04,Stojanovic++:10}.

The system Hamiltonian, describing e-ph interaction resulting 
from Rydberg dressing, can be written as
\begin{eqnarray}\label{HamExc}
H &=&\sum_n \varepsilon_n(\mathbf{u}) c_n^{\dagger}c_n +
\sum_n t_{n,n+1}(\mathbf{u})(c^\dagger_{n+1}c_n + \mathrm{H.c.}) \nonumber\\
&+& \hbar\omega_{\textrm{ph}}\sum_n b^\dagger_n b_n \:,
\end{eqnarray}
where $\mathbf{u}\equiv\{u_n|\:n=1,\ldots,N\}$ and $c^\dagger_n$ ($c_n$) creates (destroys) an excitation at site $n$; 
$\varepsilon_n (\mathbf{u})$ is the excitation on-site energy, which depends on the phonon displacements not only on site 
$n$, but also on those on the two neighboring sites $n\pm 1$. It is given by~\cite{Wuester+:11}
\begin{eqnarray} \label{vareps}
\varepsilon_n(\mathbf{u}) &=& \frac{\alpha^4\hbar\Delta}{2}\Bigg[\left\{1-\zeta^2
\left(1+\frac{u_{n+1}-u_n}{a}\right)^{-6}\right\}^{-1} \nonumber \\
&+& \left\{1-\zeta^2\left(1+\frac{u_{n}-u_{n-1}}{a}\right)^{-6}\right
\}^{-1}\Bigg]\:,
\end{eqnarray}
where the dimensionless quantity $\zeta\equiv C_3/(\hbar\Delta a^3)$ is introduced, 
with $\Delta\equiv\Delta_s+\Delta_p$. Likewise, 
\begin{eqnarray} \label{thopp}
t_{n,n+1}(\mathbf{u})&=&\frac{\alpha^4 C_3}{(a+u_{n+1}-u_n)^3}\:\nonumber\\
&\times& \left\{1-\zeta^2\left(1+\frac{u_{n+1}-u_n}{a}\right)^{-6}\right\}^{-1}
\end{eqnarray}
stands for the excitation hopping amplitude~\cite{Wuester+:11} between sites $n$ 
and $n+1$, which depends on the difference $u_{n+1}-u_n$ of the respective displacements.

\section{Effective e-ph coupling model} \label{EPHmodel}
\subsection{Single-excitation Hamiltonian with P- and B couplings} \label{SingleExcHam}
Assuming small atomic displacements (i.e., $\sqrt{\langle\hat{u}^2_n\rangle}\ll a$) it is judicious to expand 
the expressions on the right-hand-side of Eqs.~\eqref{vareps} and \eqref{thopp} to the lowest (linear) order 
in the difference of displacements. In particular, the linear dependence of $\varepsilon_n(\mathbf{u})$ on 
$u_{n+1}-u_{n-1}$ captures the antisymmetric coupling of the excitation density at site $n$ with the local phonon 
displacements on the neighboring sites $n\pm 1$ [breathing-mode-type (B) e-ph coupling]; likewise, the similar 
dependence of $t_{n,n+1}$ on $u_{n+1}-u_n$ accounts for the modulation of the hopping amplitude between 
sites $n$ and $n+1$ due to the respective phonon displacements [Peierls-type (P) coupling]~\cite{Stojanovic+Vanevic:08,
Stojanovic+:14,Stojanovic+Salom:19,StojanovicPRL:20}. To be more specific, one obtains
\begin{eqnarray}\label{onsitehopp}
\varepsilon_n(\mathbf{u}) &=& \epsilon_e +\xi_{\textrm{B}}(u_{n+1}-u_{n-1}) \:,\nonumber\\
t_{n,n+1}(\mathbf{u}) &=& -t_e + \xi_{\textrm P}(u_{n+1}-u_n) \:,
\end{eqnarray}
where the explicit expressions for $\xi_{\textrm{B}}$ and $\xi_{\textrm{P}}$ in terms of the relevant parameters of the system are
\begin{eqnarray}\label{xiBMP}
\xi_{\textrm{B}} &=& 3\:\frac{\alpha^4\hbar\Delta}{a}
\frac{\zeta^2}{\left(1-\zeta^2\right)^2}\:, \nonumber\\
\xi_{\textrm{P}} &=& 3\:\frac{\alpha^4 C_3}{a^4}
\frac{3\zeta^2-1}{\left(1-\zeta^2\right)^2}\:.
\end{eqnarray}
Likewise, the bare on-site energy $\epsilon_0$ and hopping 
amplitude $t_e$ in Eq.~\eqref{onsitehopp} are given by 
\begin{equation}\label{t_e}
\epsilon_e = \frac{\alpha^4 \hbar\Delta}{\displaystyle 1-\zeta^2}   \:, \qquad 
t_e = -\frac{\alpha^4 C_3}{a^3\displaystyle(1-\zeta^2)} \:.
\end{equation}
Importantly, the positive sign of $t_e$ [obtained for $|\zeta|>1$, i.e., 
$C_3/(\hbar|\Delta|a^3)>1$] corresponds to the usual situation where 
the bare-excitation dispersion $\epsilon(k)=\epsilon_e-2t_e\cos k$ (note that hereafter 
all quasimomenta are expressed in units of the inverse lattice period $a^{-1}$) has 
its minimum at $k=0$. The opposite, negative sign of $t_e$ [for $|\zeta|<1$, i.e., 
$C_3/(\hbar|\Delta|a^3)<1$], i.e., the case with the band minimum at $k=\pi$, 
corresponds to the bare-excitation Bloch state 
\begin{equation}\label{BareExcBloch}
|\Psi_{k=\pi}\rangle\equiv c^{\dagger}_{k=\pi}
|0\rangle_{\textrm{e}}\otimes|0\rangle_{\textrm{ph}}\:, 
\end{equation}
where $|0\rangle_{\textrm{e}}$ and 
$|0\rangle_{\textrm{ph}}$ are the respective excitation- and phonon vacuum states.

The noninteracting part of the effective system Hamiltonian consists of the free excitation 
and phonon parts:
 \begin{equation}\label{H_0}
H_{0}=\epsilon_e\sum_n c^\dagger_n c_n -t_e\sum_n (c^\dagger_{n+1}c_n 
+\mathrm{H.c.})+\hbar\omega_{\textrm{ph}}\sum_n b^\dagger_n b_n \:.
\end{equation}
The interacting part of that Hamiltonian reads $H_{\textrm{e-ph}}=H_{\textrm P}+H_{\textrm{B}}$, 
where -- with the stipulated linear approximation [cf. Eq.~\eqref{onsitehopp}] -- the P contribution 
is given by
\begin{equation}\label{H_P}
H_{\textrm P} = g_{\textrm P}\hbar\omega_{\textrm{ph}}\sum_n
(c^\dagger_{n+1}c_n+\mathrm{H.c.})(b^\dagger_{n+1} + b_{n+1} 
-b^\dagger_n - b_n)
\end{equation}
and the B part by
\begin{equation}\label{H_B}
H_{\textrm{B}} = g_{\textrm{B}}\hbar\omega_{\textrm{ph}}\sum_n 
c^\dagger_n c_n(b^\dagger_{n+1} + b_{n+1} - b^\dagger_{n-1}- b_{n-1}) \:.
\end{equation}
Here $g_{\textrm P}\equiv \xi_{\textrm{P}}/(2m\hbar\omega^{3}_{\textrm{ph}})^{1/2}$ 
and $g_{\textrm{B}}\equiv\xi_{\textrm{B}}/(2m\hbar\omega^{3}_{\textrm{ph}})^{1/2}$ 
are the respective dimensionless coupling strengths. 

The generic momentum-space form of $H_{\textrm{e-ph}}$ reads $H_{\textrm{e-ph}}=N^{-1/2}\:\sum_{k,q}
\gamma_{\textrm{e-ph}}(k,q)\:c_{k+q}^{\dagger}c_{k}(b_{-q}^{\dagger}+b_{q})$, with the e-ph vertex function 
given by
\begin{equation}\label{gamma_eph}
\gamma_{\textrm{e-ph}}(k,q)=2i\hbar\omega_{\textrm{ph}}\:\{g_{\textrm{P}}
[\sin k-\sin(k+q)]-g_{\textrm{B}}\sin q\}\:.
\end{equation}
Because this last vertex function depends not only on the excitation quasimentum $k$, but also on the 
phonon quasimomentum $q$, the total Hamiltonian of the system does 
not belong to the realm of validity of the Gerlach-L\"{o}wen theorem~\cite{Gerlach+Lowen:87}, a formal 
result that rules out a nonanalytic behavior of the ground-state energy and other relevant single-particle 
quantities (e.g., the quasiparticle residue) with varying e-ph coupling strength. In what follows it will 
be shown that the ground states of this system indeed display sharp transitions. 

For the most general (i.e. dependent on both $k$ and $q$) e-ph vertex function $\gamma_{\textrm{e-ph}}(k,q)$ 
the effective e-ph coupling strength is given by~\cite{Stojanovic:20}
\begin{equation} \label{effcouplstrength}
\lambda_{\textrm{e-ph}}=\frac{\langle|\gamma_{\textrm{e-ph}}
(k,q)|^{2}\rangle_{\textrm{BZ}}}
{2|t_{\rm e}|
\omega_{\textrm{ph}}} \:,
\end{equation}
where $\langle |\gamma_{\textrm{e-ph}}
(k,q)|^{2} \rangle_{\textrm{BZ}}$ denotes the Brillouin-zone 
average of $|\gamma_{\textrm{e-ph}}(k,q)|^{2}$ over quasimomenta $k$ and $q$: 
\begin{equation}\label{}
\langle|\gamma_{\textrm{e-ph}}(k,q)|^{2}\rangle \equiv\frac{1}{(2\pi)^2}
\:\int^{\pi}_{-\pi}\int^{\pi}_{-\pi}\:|\gamma_{\textrm{e-ph}}(k,q)|^{2}
\:dkdq  \:.
\end{equation}


\subsection{Special case: $g_{\textrm P}=g_{\textrm{B}}
\equiv g$}\label{SweetSpot}
One important special case of the effective model of the system under consideration
is the one with equal P- and B coupling strengths, i.e., $g_{\textrm P}=g_{\textrm{B}}
\equiv g$. Given that B coupling in its own right does not lead to sharp small-polaron transitions -- because its corresponding e-ph vertex function depends only on $q$ -- narrowing down the discussion of such transitions to the special case of equal P- and B coupling strengths (see Sec.~\ref{ResDisc} below) does not constitute a significant loss of generality. Another peculiarity of this special case of
the model with simultaneous P- and B couplings is that the bare-excitation Bloch state $|\Psi_{k=\pi}\rangle$ [cf. Eq.~\eqref{BareExcBloch} above]
constitutes for $t_e<0$ an exact eigenstate of the total Hamiltonian for an arbitrary coupling strength
(conversely, for $t_e>0$ the same holds for the bare-excitation Bloch state $|\Psi_{k=0}\rangle$~\cite{StojanovicPRL:20}). Moreover, below a certain critical coupling strength this state represents the ground state of the model. Therefore, the sharp transition in the special case of the model under consideration corresponds to a change from a bare (completely phonon-undressed) excitation to a strongly-dressed one (small polaron). 

This case with $g_{\textrm P}=g_{\textrm{B}}
\equiv g$ in the Rydberg-dressed neutral-atom system at hand corresponds to the ``sweet-spot'' (ss) value $\zeta_{\textrm{ss}}=(1+\sqrt{13})/6\approx 
0.77$ of the parameter $\zeta$. For chosen atomic species and the principal quantum number -- which in turn fixes the dipolar-interaction constant $C_3$ -- for an arbitrary choice of $a$ this last physical situation is realized for $\Delta^{\textrm{ss}}\equiv C_3/(\hbar\zeta_{\textrm{ss}} a^3)$;
corresponding to this last choice of the detuning is a range of Rabi frequencies $\Omega^{\textrm{ss}}=\Delta^{\textrm{ss}}\alpha$. This range is set by the adopted range of values of the dressing 
parameter $\alpha$ (see Sec.~\ref{ResDisc} below). In the following, the e-ph-coupling and total Hamiltonians of the system at hand in this special case will be denoted by 
$H^{\textrm{ss}}_{\textrm{e-ph}}$ and $H^{\textrm{ss}}$, respectively. 

To determine the e-ph coupling strength in the aforementioned special case of 
the system at hand, it is pertinent to invoke the momentum-space form of 
$H^{\textrm{ss}}_{\textrm{e-ph}}$. This momentum-space form reads 
\begin{equation}
 H^{\textrm{ss}}_{\textrm{e-ph}}=N^{-1/2}\:\sum_{k,q}
 \gamma^{\textrm{ss}}_{\textrm{e-ph}}(k,q)\:
 c_{k+q}^{\dagger}c_{k}(b_{-q}^{\dagger}+b_{q}) \:,
\end{equation}
where the e-ph vertex function $\gamma^{\textrm{ss}}_{\textrm{e-ph}}(k,q)$ 
in the last equation is given by
\begin{equation}\label{gammass}
\gamma^{\textrm{ss}}_{\textrm{e-ph}}(k,q)=2ig\hbar\omega_{\textrm{ph}}\:[\:\sin k-\sin q-\sin(k+q)]\:.
\end{equation}
As a result, the effective e-ph coupling strength
[cf. Eq.~\eqref{effcouplstrength} above] in the problem 
at hand evaluates to 
$\lambda^{\textrm{ss}}_{\textrm{e-ph}}
\equiv 3g^{2}\:\hbar\omega_{\textrm{ph}}/|t_{\rm e}|$.
In terms of the relevant parameters of the system under
consideration, this effective coupling strength is given by
\begin{equation}\label{lambda_param}
\lambda^{\textrm{ss}}_{\textrm{e-ph}}=\frac{27}{2}\:\alpha^4\:\frac{C_3}
{m\omega_{\textrm{ph}}^{2} a^5}\:\frac{(3\zeta_{\textrm{ss}}^2-1)^2}
{(1-\zeta_{\textrm{ss}}^2)^3}\:.
\end{equation}
The resulting dependence of $\lambda^{\textrm{ss}}_{\textrm{e-ph}}$ 
on $\alpha\equiv\Omega^{\textrm{ss}}/\Delta^{\textrm{ss}}$ leads to 
the conclusion that the Rabi frequency $\Omega^{\textrm{ss}}$ is the 
principal experimental knob in the system under consideration. Therefore,
by varying this last quantity different characteristic regimes of 
the system can be accessed.

\section{Results and Discussion} \label{ResDisc}
In what follows, the proposed scheme for analog quantum simulation of small-polaron physics is discussed for the value $n_{\textrm q}=80$ of the principal quantum number; the corresponding value of the dipolar-interaction constant $C_3$ for $^{87}$Rb atoms is $2\pi\hbar\times 40$\:GHz$\mu$m$^3$. 

While the system under consideration can accommodate 
two different regimes as far as the sign of the effective bare-excitation hopping amplitude $t_e$ is concerned [cf. Sec.~\ref{SingleExcHam}], the following discussion will be confined to the case where $t_e<0$ [ realized for $|\zeta|<1$ (i.e., $C_3/(\hbar|\Delta|a^3)<1$) ], i.e. the case with the 
bare-excitation band minimum corresponds to $k=\pi$. The case with $t_e>0$ 
can be discussed in an analogous fashion.

To facilitate further analysis, it is pertinent to fix at this point realistic 
ranges of values for the relevant parameters of the system. As usual for neutral atoms in 
optical-tweezer arrays, the period $a$ varies between around $3\:\mu$m 
and tens of micrometers. Accordingly, the attendant values of the 
ss detuning can vary in an extremely wide range: e.g., for $a=4$\:$\mu$m 
one obtains $\Delta^{\textrm{ss}}\approx 5.12$\:GHz, while for 
$a=15$\:$\mu$m one has $\Delta^{\textrm{ss}}\approx 97$\:MHz. In 
addition, the typical values for the trapping frequency are 
$\omega_{\textrm{ph}}/(2\pi)\sim(2 - 5)$\:kHz. Finally, for the 
dressing parameter it is pertinent to consider the range 
$\alpha = 0.01 - 0.1$.

While the entire phase diagram of the system under consideration could be
obtained using the same methodology, the discussion that follows will be 
restricted to the special case with equal P- and B coupling strengths 
[recall the discussion in Sec.~\ref{SweetSpot} above]. This 
special case of the effective model of the system at hand 
is particularly interesting as the corresponding sharp small-polaron transition -- obtained for a critical value
of the effective e-ph coupling strength (i.e., for a critical value of 
the experimental knob -- the Rabi frequency) -- in that case signifies 
the change from a completely undressed (bare) itinerant excitation 
(below the critical value of the experimental knob, i.e. below 
the critical effective e-ph coupling strength) to a strongly 
phonon-dressed one (above the critical value of the experimental knob).

The ground-state energy of the system can be calculated using 
Lanczos-type exact diagonalization on properly truncated Hilbert 
space of the coupled e-ph system~\cite{Stojanovic:20}. To be more 
specific, exact diagonalization of the Hamiltonian $H^{\textrm{ss}}=H_0+H^{\textrm{ss}}_{\textrm{e-ph}}$ is performed here for a finite 
system with $N=10$ sites (i.e., atoms) and the maximal number $M=8$ 
of phonons in the truncated phonon Hilbert space. This is done using 
the well-established approach for truncating bosonic Hilbert spaces in a controllable fashion.
This approach rests on a gradual increase of $N$ -- with a simultaneous 
increase of $M$ -- until a numerical convergence of the obtained 
ground-state energy and other ground-state-related quantities is achieved~\cite{Stojanovic:20}. 

\begin{figure}[t!]
\includegraphics[clip,width=8.25cm]{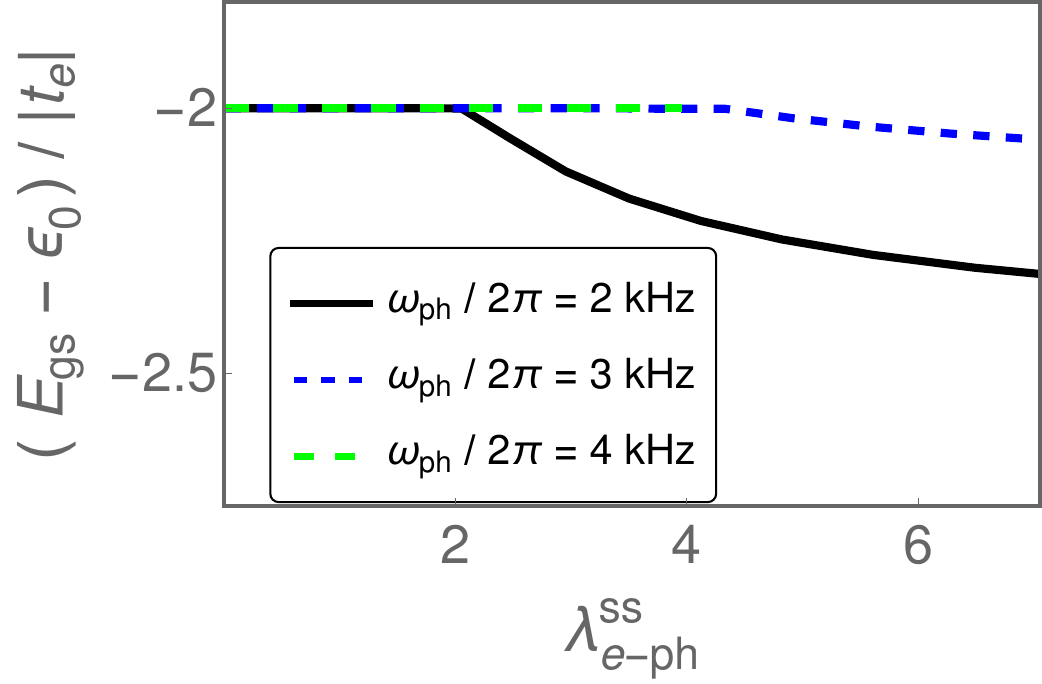}
\caption{\label{fig:EnergyVsLambda} Ground-state energy of the system
(without the on-site-energy contribution $\epsilon_0$), shown as a function 
of the effective coupling strength $\lambda^{ss}_{\textrm{e-ph}}$. These results
correspond to the value $a = 4\:\mu$m of the period of tweezer array and three different values of the phonon frequency $\omega_{\textrm{e-ph}}$.}
\end{figure}

\begin{figure}[b!]
\includegraphics[clip,width=8.25cm]{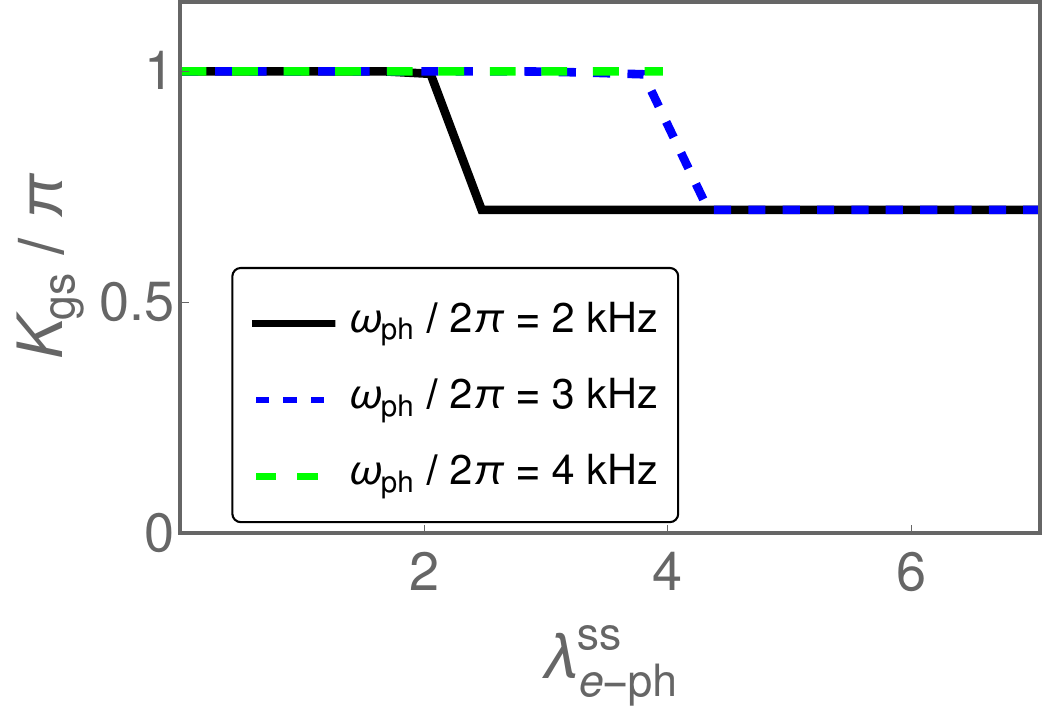}
\caption{\label{fig:GSmomentumVsLambda} Ground-state quasimomentum of the system, i.e. quasimomentum corresponding to the energy minumum, shown as a function 
of the effective coupling strength $\lambda^{ss}_{\textrm{e-ph}}$. These results correspond to $a = 4\:\mu$m and three different values of the phonon frequency $\omega_{\textrm{e-ph}}$.}
\end{figure}

The aforementioned numerical evaluation demonstrates that the ground state 
of $H^{\textrm{ss}}$ undergoes a sharp, level-crossing transition~\cite{Stojanovic:20} at a certain critical value $(\lambda^{\textrm{ss}}_{\textrm{e-ph}})_\textrm{c}$ of $\lambda^{\textrm{ss}}_{\textrm{e-ph}}$.  
To be more specific, the ground state of this Hamiltonian
for $\lambda^{\textrm{ss}}_{\textrm{e-ph}}<(\lambda^{\textrm{ss}}_{\textrm
{e-ph}})_\textrm{c}$ is at the same time the eigenvalue $\pi$ of $K_{\mathrm{tot}}$ 
and has a rather peculiar character. Namely, despite being the ground state 
of an interacting e-ph Hamiltonian, this state coincides with the bare-excitation Bloch state $|\Psi_{k=\pi}\rangle$ [cf. Eq.~\eqref{BareExcBloch} above]; accordingly,
its corresponding energy $\epsilon_0-2|t_{e}|$ is the minimum of a 1D 
tight-binding dispersion. On the other hand, the ground state of the system 
for $\lambda^{\textrm{ss}}_{\textrm{e-ph}}\ge (\lambda^{\textrm{ss}}_{\textrm{e-ph}})_\textrm{c}$ is strongly phonon-dressed, i.e. has the character of 
a small polaron. This polaronic ground state is also twofold-degenerate 
and corresponds to $K=\pm K_{\textrm{gs}}$, where $0<K_{\textrm{gs}}<\pi$. 
The obtained results for ground-state energy $E_{\textrm{gs}}$ of the system -- expressed in units of $|t_{e}|$ -- on the effective e-ph coupling strength
are displayed in Fig.~\ref{fig:EnergyVsLambda}, in which the constant contribution $\epsilon_0$ is omitted.

The dependence of the ground-state total quasimomentum $K_{\textrm{gs}}$ 
on the effective coupling strength $\lambda^{\textrm{ss}}_{\textrm{e-ph}}$
is illustrated in Fig.~\ref{fig:GSmomentumVsLambda}. In keeping with the aforementioned bare-excitation character of the ground state below the critical values of the effective coupling strength, the total quasimomentum $K_{\textrm{gs}}=\pi$ has a vanishing phonon contribution as $\langle\Psi_{k=\pi}|\sum_n b^\dagger_n b_n|\Psi_{k=\pi}\rangle=0$. 
The circumstance that this bare-excitation Bloch state 
$|\Psi_{k=\pi}\rangle$ is the 
ground-state of a coupled e-ph
model directly follows from the assumption that the two relevant 
e-ph coupling strengths are equal to one another (i.e. $g_{\textrm P}=
g_{\textrm{B}}\equiv g$; recall Sec.~\ref{SweetSpot}). In other words, 
this is a direct implication of an effective mutual cancellation of 
P- and B couplings for a bare excitation with quasimomentum $k=\pi$.

\begin{figure}[t!]
\includegraphics[clip,width=8.25cm]{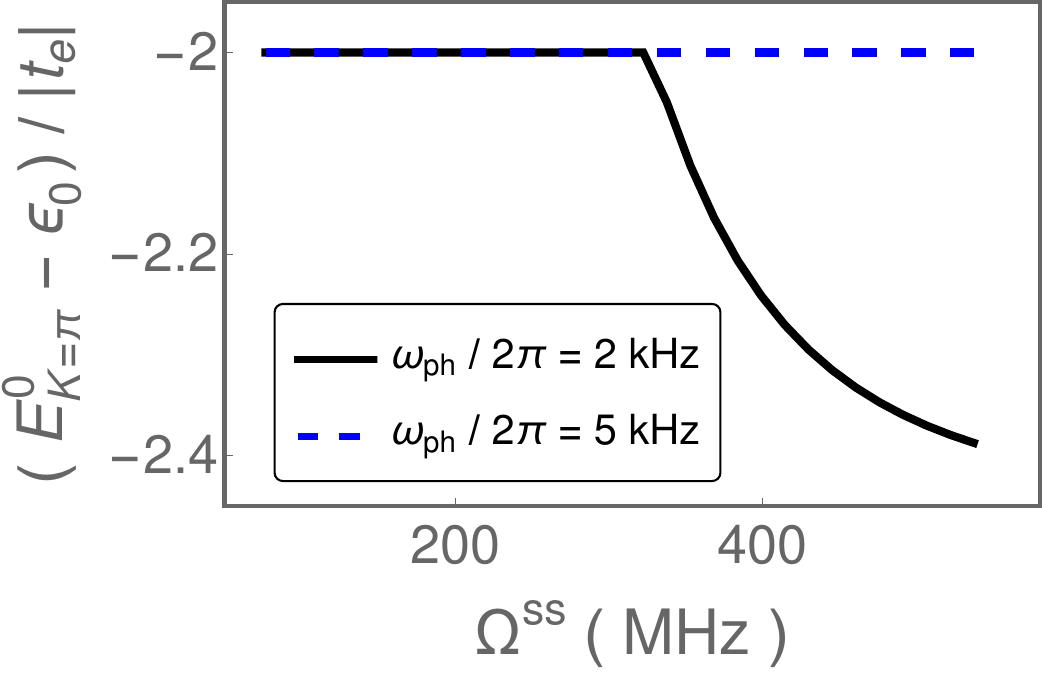}
\caption{\label{fig:EnergyVsOmega} Illustration of the sharp small-polaron transition: the $K=\pi$ eigenvalue (without the constant energy offset $\epsilon_0$) 
of the total Hamiltonian $H^{\textrm{ss}}=H_0+H^{\textrm{ss}}_{\textrm{e-ph}}$ of the system as a function of the Rabi frequency $\Omega^{\textrm{ss}}$ for two different choices of
the phonon frequency $\omega_{\textrm{ph}}$ and 
$a=4\:\mu$m.}
\end{figure}

Given that the Rabi frequency $\Omega^{\textrm{ss}}$
represents the main experimental knob in the neutral-atom system under consideration, it is of interest to illustrate the exact location of the first sharp small-polaron transition in this system as a function of this parameter. Such an illustration is provided in Fig.~\ref{fig:EnergyVsOmega}, which shows the 
the $K=\pi$ eigenvalue (without the constant energy offset $\epsilon_0$) 
of the total Hamiltonian $H^{\textrm{ss}}=H_0+H^{\textrm{ss}}_{\textrm{e-ph}}$ of the system as a function of 
$\Omega^{\textrm{ss}}$ for two different choices of the phonon frequency 
$\omega_{\textrm{ph}}$. For $\omega_{\textrm{ph}}=2\pi\times5\:$KHz (the dashed line) the system does not undergo a sharp transition as the effective coupling is below the critical value in the entire range of Rabi frequencies shown; therefore, in this case the lowest-energy state of the system 
coincides with the bare-excitation Bloch state $|\Psi_{k=\pi}\rangle$ (its energy corresponds to 
the band minimum for a bare excitation). On the other hand, for $\omega_{\textrm{ph}}=2\pi\times 5\:$KHz (the solid line) a sharp small-polaron transition takes place for $\Omega^{\textrm{ss}}\sim 330\:$MHz. 
Below this critical value the ground state of the system coincides to a bare (completely undressed by phonons) excitation with quasimomentum $K=k=\pi$, while above this value the ground state is 
heavily phonon-dressed, i.e. it acquires the character of a small polaron. 

Needless to say, the sharp transition illustrated in Fig.~\ref{fig:EnergyVsOmega} is just the initial one in an entire sequence of sharp transitions. Namely, in keeping with the already established phenomenology of sharp small-polaron transitions, as the value of the experimental knob $\Omega^{\textrm{ss}}$ is varied above the first sharp-transition 
point, a sequence of transitions takes place that accommodates the gradual change of the ground-state quasimomentum $K_{\textrm{gs}}$ from $K_{\textrm{gs}}=\pi$ [note that in the conventional case with a positive bare-excitation hopping amplitude ($t_e>0$) one has $K_{\textrm{gs}}=0$ instead] to the final value $K_{\textrm{gs}}=\pi/2$ that is achieved deeply in the strong e-ph coupling regime. 

\section{Summary and Conclusions} \label{SummConcl}
To summarize, in this paper it was shown that the phenomenology of sharp
small-polaron transitions can be demonstrated within the framework of an 
analog quantum simulator based on an array of neutral atoms in optical tweezers,
which are assumed to interact through Rydberg-dressed resonant dipole-dipole
interaction. In particular, it was demonstrated here that the effective model
of this system describes an itinerant excitation interacting with quanta of 
nearly-harmonic vibrations -- mimicking Einstein-type (i.e. zero-dimensional)
phonons -- through two different nonlocal excitation-phonon coupling mechanisms
(Peierls- and breathing-mode-type e-ph interactions) of interest in realistic 
electronic materials. Furthermore, it was shown that with experimentally feasible 
values of the relevant parameters of the envisioned system (e.g., the period of 
the tweezer array, detunings of the external dressing lasers from the internal
transitions, etc.) one can reach strong enough excitation-phonon coupling to
demonstrate a sharp polaronic transition. To facilitate future experimental 
realizations, such a transition was discussed here in the special case with 
equal Peierls- and breathing-mode coupling strengths; this special case is 
particularly illustrative as the corresponding sharp transition marks the change
from a completely bare (completely undressed by phonons) excitation (below the 
sharp-transition point) a strongly phonon-dressed one (above the transition point).

The demonstrated capability of the envisioned neutral-atom-based analog quantum 
simulator to facilitate the realization of a sharp small-polaron transition is 
particularly valuable as typical electron-phonon coupling strengths in realistic electronic materials 
are not sufficiently large for observing such sharp transitions. While the present work was 
solely concerned with the ground-state properties of small polarons resulting from strong 
nonlocal (strongly momentum-dependent) interactions with dispersionless phonons,
the proposed analog simulator could also be employed to study the nonequilibrium 
dynamics of small-polaron formation following a quench of excitation-phonon interacation.
For these reasons, an experimental realization of this system with neutral atoms in 
arrays of optical tweezers is keenly anticipated.

\end{document}